\documentclass[runningheads]{llncs}
\pdfoutput=1
\usepackage{cite}
\usepackage[normalem]{ulem}
\usepackage{makecell}
\usepackage{amsmath,amssymb,amsfonts}
\usepackage{algorithmic}
\usepackage{graphicx}
\usepackage{multirow}
\usepackage{textcomp}
\usepackage{bbding}
\usepackage{xcolor}
\usepackage{array} 
\usepackage{xspace}
\usepackage{float}
\usepackage{rotating} 
\usepackage[misc]{ifsym}
\useunder{\uline}{\ul}{}

\begin{document}

\newcommand{\modelacro}{\texttt{GENET}\xspace}
\newcommand{\modelfull}{\texttt{Generalized hypErgraph pretraiNing on sidE informaTion}\xspace}
\newcommand{\ourmodel}{\text{GENET}}
\newcommand{\pretext}{\text{LP-PM}}
\newcommand{\hgselfcl}{\text{HSCL}}

\newcommand{\hg}{\mathcal{G}}
\newcommand{\userset}{\mathcal{U}}
\newcommand{\itemset}{\mathcal{V}}
\newcommand{\usernode}{u}
\newcommand{\itemnode}{v}
\newcommand{\lhgconv}{\mathbf{LHGConv}}
\newcommand{\hyperedge}{e}
\newcommand{\hgmatrix}{\mathbf{H}}
\newcommand{\nodeset}{\mathcal{X}}
\newcommand{\hyperedgeset}{\mathcal{E}}
\newcommand{\node}{x}
\newcommand{\brand}{b}
\newcommand{\region}{g}
\newcommand{\cate}{c}
\newcommand{\product}{p}
\newcommand{\regionnum}{r}
\newcommand{\posreviewrate}{\rho}

\newcommand{\nodematrix}{\mathbf{X}}
\newcommand{\hyperedgematrix}{\mathbf{E}}

\newcommand{\hgencoder}{\mathbf{LHGE}}
\newcommand{\nodepert}{\text{NP}}
\newcommand{\incimatrixpert}{\text{IMP}}
\newcommand{\uigraph}{\mathcal{A}}
\newcommand{\embnode}{\mathbf{x}}
\newcommand{\embhedge}{\mathbf{e}}
\newcommand{\embuser}{\mathbf{u}}
\newcommand{\embitem}{\mathbf{v}}
\newcommand{\ftembuser}{\overline{\mathbf{u}}}
\newcommand{\ftembitem}{\overline{\mathbf{v}}}

\newcommand{\ploss}{\mathcal{L}^{P}}
\newcommand{\ftloss}{\mathcal{L}^{F}}
\newcommand{\preloss}{\mathcal{L}^{Pre}}
\newcommand{\clloss}{\mathcal{L}^{C}}
\newcommand{\overalltopnloss}{\mathcal{L}^{Top}}
\newcommand{\overallseqloss}{\mathcal{L}^{Seq}}
\newcommand{\intraloss}{\mathcal{L}^{intra}}
\newcommand{\interloss}{\mathcal{L}^{inter}}

\newcommand{\initnodes}{\mathbf{X^0}}

\newcommand{\embnodes}{\mathbf{Z^n}}
\newcommand{\embhedges}{\mathbf{Z^e}}

\newcommand{\finetnode}{\mathbf{f}}
\newcommand{\itemseq}{\mathcal{I}}
\newcommand{\seqloss}{\mathcal{L}^S}

\newcommand{\nodes}{x}
\newcommand{\hedges}{e}

\title{GENET: Unleashing the Power of Side Information
for Recommendation via Hypergraph Pre-training}

\titlerunning{GENET}
%
%
%

\author{Yang Li\inst{1,5}\orcidID{0000-0001-8501-1814} \and
Qi'ao Zhao\inst{2}\orcidID{0009-0008-0137-1934} \and
Chen Lin\Letter\inst{2,5}
\orcidID{0000-0002-2275-997X}
\and
Zhenjie Zhang\inst{3}\orcidID{0000-0001-7242-8215} \and
Xiaomin Zhu\inst{4}\orcidID{0000-0003-1301-7840} \and
Jinsong Su\inst{2,5}\orcidID{0000-0001-5606-7122} 
}
\authorrunning{Y. Li et al.}

\institute{Institute of Artificial Intelligence
, Xiamen University, Xiamen, China \email{goatxy@stu.xmu.edu.cn}\and
School of Informatics
, Xiamen University, Xiamen, China \email{zhaoqiao@stu.xmu.edu.cn}\\
\email{\{chenlin,jssu\}@xmu.edu.cn}
\and Neuron Mobility Pte.Ltd, Singapore
\email{zhenjie.zhang@neuron.sg}\\
 \and
Strategic Assessments and Consultation Institute
, Academy of Military Science
Beijing, China\\
\email{xmzhu@nudt.edu.cn}
\and Shanghai Artificial Intelligence Laboratory, Shanghai, China}

\maketitle              
\begin{abstract}
Integrating side information in recommendation systems to address user feedback sparsity has gained significant research interest. However, existing models face challenges in generalization across different domains and types of side information. Specifically, three unresolved challenges are (1) the diverse formats of side information, including text sequences and numerical features; (2) the varied semantics of side information that describes users and items at multiple levels; (3) the challenge of measuring diverse correlations in side information beyond pairwise relationships. In this paper, we introduce GENET (Generalized hypErgraph pretraiNing on sidE informaTion), that pre-trains user and item representations on feedback-irrelevant side information and fine-tunes the representations on user feedback data. GENET utilizes pre-training to prevent side information from overshadowing critical feedback signals. It employs a hypergraph framework to accommodate various types of diverse side information. During pre-training, GENET integrates tasks for hyperlink prediction and self-supervised contrast to capture fine-grained semantics at both local and global levels. Moreover, it introduces a unique strategy to enhance pre-training robustness by perturbing positive samples while maintaining high-order relations. Extensive experiments demonstrate that GENET exhibits strong generalization capabilities, outperforming the SOTA method by up to $38\%$ in TOP-N recommendation and sequential recommendation on various datasets. 
\end{abstract}
\begin{figure*}[!htbp]
	\centering
	\includegraphics[width=1\linewidth]{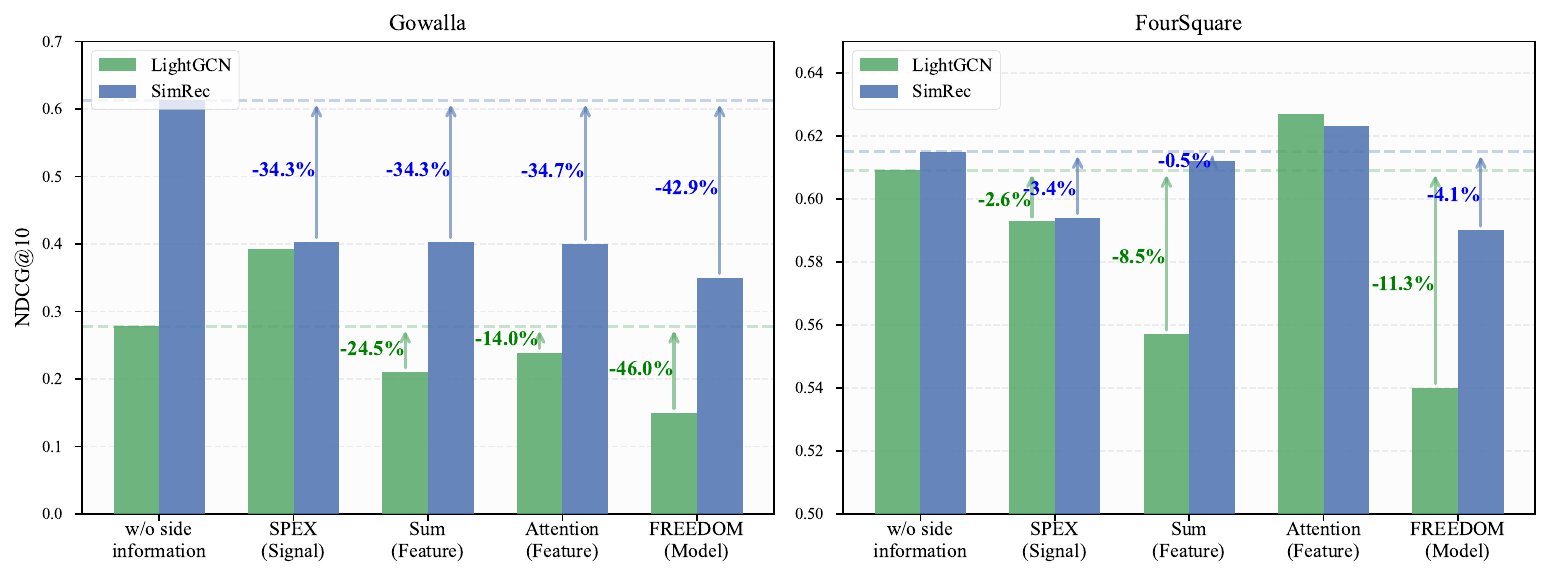}
	\caption{NDCG@10 on the Gowalla and FourSquare dataset by incorporating side information at feature, model, and signal level upon backbone models LightGCN and SimRec}
	\label{fig:intro}
\end{figure*}
\section{INTRODUCTION}
\label{sec:intro}
A major and ongoing thrust of research on Recommendation Systems (RSs) is to leverage side information. 
Side information, a.k.a, auxiliary data, which refers to non-feedback data associated with users and items~\cite{fang2011matrix,rao2022graph,liu2019recommender}, is available in various forms. 
For example, most RSs provide rich demographic profiles of users~\cite{zhou2020s3,liu2019recommender}, e.g., gender, location, etc. 
Numerous item features~\cite{xie2022decoupled,zhou2022tale,fang2011matrix} can be accumulated on an E-commerce RS, including the item's category, brand, price, textual descriptions, display images, and so on. 
Users of RSs also generate a wealth of data on item reviews~\cite{zhao2016predictive,ning2012sparse}. 
Due to the sparsity of user feedback, side information is valuable for enhancing recommendation~\cite{liu2019recommender,pfadler2020billion}, especially on cold-start users and items~\cite{pfadler2020billion}.

Existing methods of utilizing side information can be roughly classified into three types: 
(1) extract and combine the \emph{feature} from side information by direct addition~\cite{wang2018billion} or an attention mechanism~\cite{fang2011matrix,xie2022decoupled}; 
(2) utilize an additional \emph{model}, e.g., a graph~\cite{zhou2022tale} or a hypergraph~\cite{YuYLWH021MHCN}, to capture user-item relationships; 
(3) incorporate side information as a source of supervision \emph{signal}, e.g., multi-task learning~\cite{li2021spex}.

However, \emph{existing methods often fail to generalize well to different domains and/or side information}. 
To demonstrate this, we select four recent studies that incorporate side information at feature (i.e., directly summing features~\cite{wang2018billion} and combining features with attention ~\cite{zhang2019feature}), model (i.e., FREEDOM\footnote{We create an item-item graph for FREEDOM based on the similarity of POI, and we adapted the FREEDOM implementation to the SimRec framework.}~\cite{zhou2022tale}), and signal (i.e., SPEX~\cite{li2021spex}) level, and apply them on two backbones, including the commonly adopted LightGCN~\cite{he2020lightgcn} and the state-of-the-art SimRec~\cite{xia2023graph}. 
Then, we implement these methods to perform POI recommendations by utilizing social network information on the Gowalla~\cite{yin2015modeling} and FourSquare~\cite{feng2018deepmove} datasets. 
As shown in Figure \ref{fig:intro}, the inclusion of various side information in most cases reduces the performance of the backbone model.  
For example, FREEDOM was designed to utilize item textual and visual descriptions and had gained a $15.65\%$ increase over LightGCN on Amazon Sports dataset as per~\cite{zhou2022tale}. But as shown in Figure~\ref{fig:intro}, FREEDOM still can't match SimRec without side
information, and even using SimRec as its backbone, FREEDOM is still inferior to SimRec itself. 
SPEX was designed to utilize social network information to improve social recommendation and had gained a $13.48\%$ increase on the Weibo dataset as reported in~\cite{li2021spex}. Unfortunately, it has led to a significant performance decrease in backbone SimRec. 
Furthermore, existing methods with side information are inferior to SimRec, which is based solely on feedback data without side information.
Therefore, \emph{it is critical to develop a generalized model that can utilize various side information to enhance recommendation performance across domains}.  

We argue that the malfunction of existing methods is primarily due to the direct involvement of voluminous side information in recommendation learning, often exceeding the quantity of user/item data and feedback. This can lead to the overshadowing of essential ID features and imbalance in models that treat side information separately, degrading performance. Additionally, the indirect nature of certain side information, like social networks, and the risk of negative transfer from noisy supervisory signals, further complicate their integration in recommendation systems. \emph{Pre-training-fine-tuning} may alleviate these problems. Its paradigm first learns representations from raw data in a self-supervision manner and then fine-tunes the representations on the downstream tasks. 
The goal is to benefit target tasks with knowledge acquired in pre-training and mitigate negative transfer.

In this paper, we propose to learn user- and item-representations by \emph{pre-training on the side information} and \emph{fine-tuning on the feedback data}. 
We focus on solving three challenges with regard to the generalization of pre-training on side information. 

\textbf{C1: diversity of format}. 
Previous pre-training methods are based on either sequence models~\cite{devlin-etal-2019-bert} or graph neural networks~\cite{hu2020gpt}, and they are natural for modeling one format but sub-optimal for others. 
\textbf{C2: diversity of semantics}. Side information describes users and items from multi-level in a context different from RSs.  
Prior studies often tailor pre-training tasks to different side information, e.g., link prediction task in the social network learns node semantics from a global perspective of user collaboration~\cite{el2022twhin}, contrastive learning tasks with domain-specific data augmentations~\cite{dong2022m5product} learns item semantics from a local perspective of textual descriptions, and these pre-training tasks are not robust across domains.  
\textbf{C3: diversity of correlation}. Side information measures similarity/dissimilarity over pairs, triples, or more objects, and these multi-faceted and/or high-order relationships are important. 
A former study emphasizes pair-wise relationships~\cite{li2013recommendation} and overlooks the high-order and multi-faceted relationship. 

In this paper, we propose \modelacro, meaning Generalized hypergraph pretraining on sidE informaTion.
To address \textbf{C1}, \modelacro is based on a hypergraph suitable for representing heterogeneous side information. 
To address \textbf{C2}, \modelacro presents three pre-training tasks, i.e., hyperlink prediction, global contrastive, and local contrastive, to reveal semantic relationships among users/items at different levels and combine fine-grained and coarse-grained representations. 
To address \textbf{C3}, \modelacro presents a novel strategy to corrupt the positive sample in the hyperlink prediction task to increase the robustness of pertaining tasks while preserving the high-order and multi-faceted relation. 


In summary, the main contributions are three-fold. (1) A unified model \modelacro is presented to unleash the power of side information and enhance recommendation performance across different domains. (2) Two pre-training tasks based on hypergraph properties, namely hyperlink prediction (GENET-P) and hypergraph contrastive learning (HSCL), are introduced to better explore high-order and multi-faceted relationships. (3) Extensive experiments (i.e., on three different real-world datasets, two recommendation tasks, and many side information) demonstrate that the proposed \modelacro significantly outperforms the SOTA competitor up to 38\% improvement in TOP-N recommendation and Sequential recommendation tasks and show the powerful potential in cold-start recommendations by at least $175\%$ improvement on all datasets. Our code is available online.\footnote{https://anonymous.4open.science/r/GENET-D582/}

\section{RELATED WORK}
We briefly introduce relevant studies on recommendation models with side information, pre-training, and hypergraphs. 

\subsection{Recommendation Models with Side Information} \label{sec:related}
Using side information to improve recommendation has been extensively studied~\cite{fang2011matrix,zhou2022tale,li2021spex,rao2022graph,yang2020location}. 
This section discusses three methods of improving recommendation systems using side information. The first method involves combining features from side information, exemplified by HIRE~\cite{liu2019recommender} and FDSA~\cite{fang2011matrix} which integrate these details in different ways. The second approach employs additional models, like graph models, to capture relationships between users and items, with FREEDOM~\cite{zhou2022tale} and Flashback~\cite{yang2020location} constructing various item graphs. The third method uses side information as a supervisory signal, as seen in SPEX~\cite{li2021spex} which incorporates social network data through multitask learning. Although these methods show improvements, they may fall short in representing heterogeneous side information and in generalization capabilities. 

\subsection{Recommendation Models with Pre-training}
Pre-training in recommendation models is mainly divided into two categories: one based on user-item feedback data and the other on side information. In the first category, models like SimRec~\cite{xia2023graph} and BERT4Rec~\cite{sun2019bert4rec} enhance robustness and efficiency by pre-training on user-item feedback. The second category, exemplified by Graph-Flashback~\cite{rao2022graph} and S$^3$-rec~\cite{zhou2020s3}, employs various pre-training strategies to learn representations from side information, enriching user and item profiles. 
These methods overlook the diversity of data formats and lose high-order information.

\subsection{Recommendation models based on Hypergraph}
Recent studies increasingly focus on using hypergraphs for recommendations. For instance, HyperRec\cite{WangDH0C20HyperRec} employ hypergraph convolution networks to capture short-term correlations in sequence recommendation tasks and combine prediction tasks with contrastive tasks on hypergraphs. MHCN~\cite{YuYLWH021MHCN} introduces a multi-channel hypergraph convolutional network to enhance social recommendations. We identify a potential risk in previous methods: the direct use of side information in training recommendation tasks, leading to negative transfer effects. 

Our method leverages pre-trained representations on a hypergraph, integrating diverse side information like sequences and graphs. This strategy alleviates the above issues, may prevent negative transfer effects, and offers a versatile solution for different side information types. 
Recently, we discovered another work on hypergraph pre-training that has been published as a preprint\cite{yang2023unified}. We are sorry that we may not compare with it due to the lack of finalized content and code. In contrast, our framework is a unified approach that encompasses various types of side information, specifically designed pre-training tasks for hypergraph structure, and includes evaluations across multiple downstream tasks.

\section{Pre-training on Side Information}
To pre-train on side information, we first construct a hypergraph on the side information (Section~\ref{sec:construct}), perform propagation on the hypergraph to obtain node embeddings for users and items (Section~\ref{sec:convolution}) and optimize the node embeddings via a set of pre-training tasks (Section~\ref{sec:linktask} to Section~\ref{sec:contrasttask}).

\begin{figure}[!t]
	\centering
	\includegraphics[width=1\linewidth]{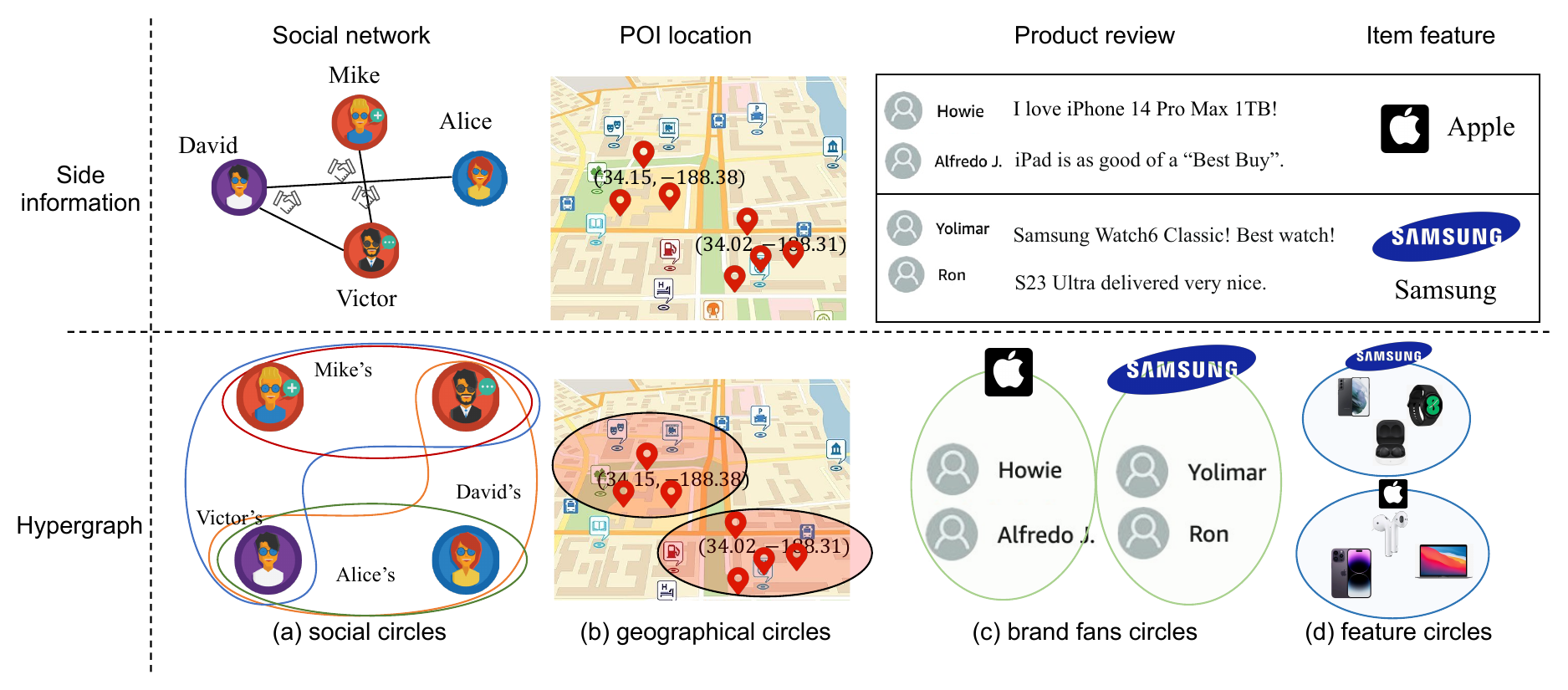}
	\caption{Different types of side information and the construction of hypergraphs}
	\label{fig:sideinfo}
\end{figure}
Formally, a hypergraph $\hg=(\nodeset,\hyperedgeset)$ consists of a set of vertices $\nodeset$ and a set of hyperedges $\hyperedgeset$, where $\nodeset=\{\node_1,\node_2,\cdots,\node_{|\nodeset|}\}$, each node can be either a user or an item.  $\hyperedge=\{\hyperedge_1,\hyperedge_2,\cdots,\hyperedge_{|\hyperedgeset|}\}$, each edge connects multiple nodes. A node $\node_i$ and a hyperedge $\hyperedge_p$ are incident if $\node_i\in \hyperedge_p$. The incidence matrix 
$\hgmatrix \in \{0,1\}^{|\nodeset|\times|\hyperedgeset|}$ describes the structure of $\hg$, where $h(\node_i,\hyperedge_p)=1$ if $\node_i$ and $\hyperedge_p$ are incident. 
Two nodes $\node_i,\node_j$ are adjacent if there exists at least one hyperedge $\hyperedge_p$ that $\node_i\in\hyperedge_p,\node_j\in\hyperedge_p$. 

\subsection{Construction of Hypergraphs}\label{sec:construct} 
In constructing hypergraphs ($\hg$), we focus on four types of side information: social networks, POI locations, product reviews, and item brands. These encompass various data formats such as relational data, categorical and numerical features, and text sequences, suitable for common recommendation scenarios like POI and e-commerce recommendations.
\begin{itemize}
\item \textbf{Social Network Hypergraph:} Constructs hyperedges to represent social circles, capturing high-order and multifaceted social connections. For example, a hyperedge represents David's social circle, connecting individuals like Victor and Alice if they are friends of David. It is shown in Figure~\ref{fig:sideinfo}(a).

\item \textbf{POI Location Hypergraph:} Utilizes geographical data (longitude and latitude) to segment POIs into regions. POIs are clustered into regions using k-means clustering, and a hyperedge is constructed to represent each region. It is shown in Figure~\ref{fig:sideinfo}(b).

\item \textbf{Product Review Hypergraph:} Construct a hyperedge of the fan circle for each product brand. We mine the sentiment of user reviews for branded products, identifying those users who consistently give high ratings to the brand's products, and define them as 'fans' of the brand, shown in Figure~\ref{fig:sideinfo}(c).

\item \textbf{Item Feature Hypergraph:} Handles item-related metadata, such as brands and categories. Hyperedges are constructed for each brand and category. An item is connected to the corresponding hyperedges if it belongs to a specific brand or category. It is shown in Figure~\ref{fig:sideinfo}(d).
\end{itemize}

\subsection{Node Representation via Hypergraph Convolution}\label{sec:convolution}
\textbf{Input} of $\modelacro$ is a hypergraph $\hg$, a sparse matrix of the initial node embeddings $\nodematrix^0 \in \mathbb{R}^{|\nodeset|\times|\nodeset|}$, where each node is represented by one-hot encoding, the degree matrix of hyperedges $\mathbf{D}^{e}$ and the degree matrix of nodes$\mathbf{D}^{v}$.

\begin{figure*}[t]
	\centering
	\includegraphics[width=1\linewidth]{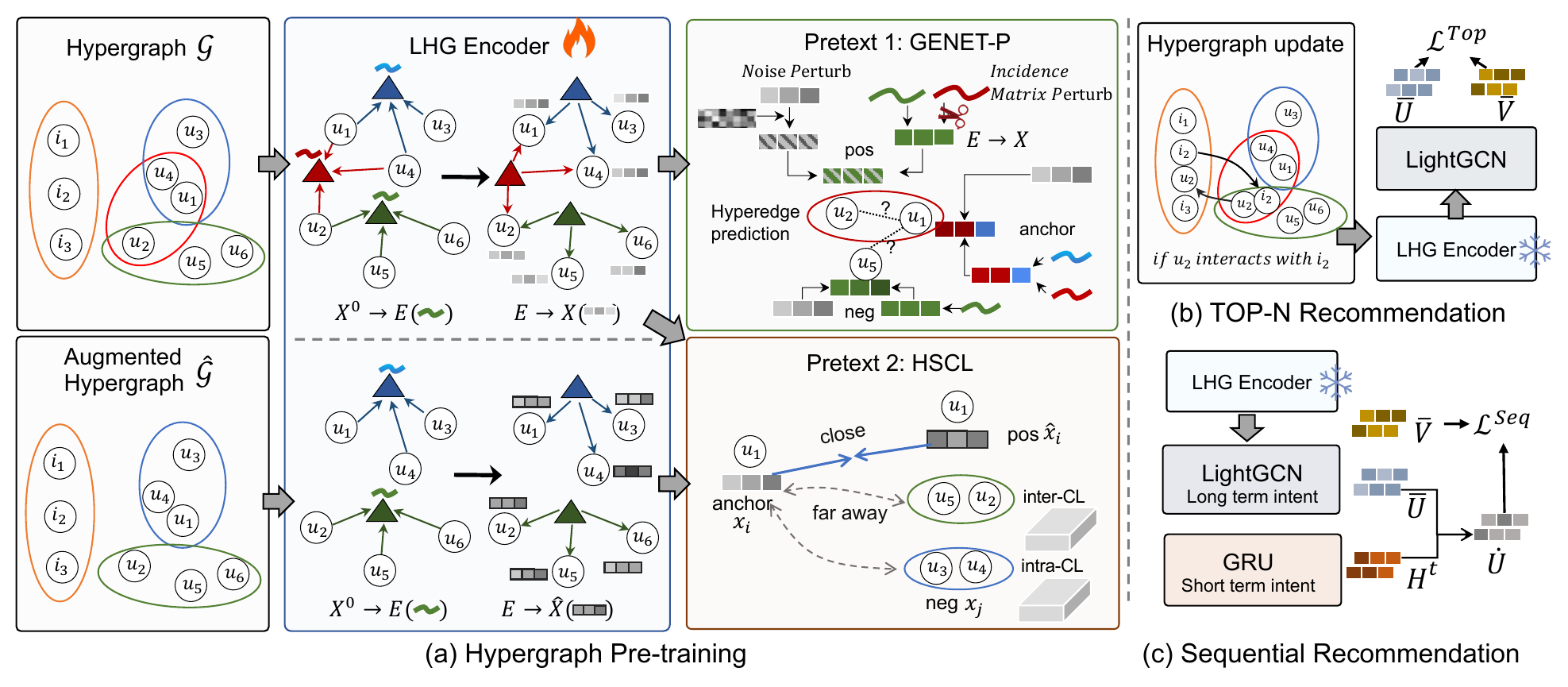}
	\caption{The framework of $\modelacro$ pre-training phase and fine-tuning stage.}
	\label{fig:framework}
\end{figure*}

\textbf{LHG Encoder} Inspired by the general Hypergraph Neural Network HGNN$^+$\cite{gao2022hgnn}, we derive the node embeddings by introducing an edge embedding vector for each hyperedge and propagating edge embeddings to incident nodes. As shown in Figure~\ref{fig:framework}(a),  the propagation is performed from nodes to hyperedges and from hyperedges to nodes, as shown below:
\begin{equation}
\begin{aligned}
\hyperedgematrix &= 
\mathbf{D}^{e^{-1}}\mathbf{H}^{\top}\nodematrix^{0}\mathbf{\Theta}^{0}, \\
\nodematrix &= \mathbf{D}^{v^{-1}}\mathbf{H}\mathbf{I}\hyperedgematrix.
\end{aligned}
\end{equation}
where $\hyperedgematrix$ is the edges embedding matrix. $\mathbf{\Theta}$ is the trainable matrix. $\nodematrix$ is the nodes embedding matrix. $\mathbf{I}$ is an identity matrix. Noted that our encoder is Light HyperGraph encoding (LHG), because different from HGNN${^+}$, we remove the non-linear activation function, which may affect the aggregation of information in the hypergraph.

\subsection{Hyperlink Prediction}\label{sec:linktask}
One pre-training task on $\hg$ is hyperlink prediction, i.e., to predict if two nodes are adjacent given a hyperedge. 
The motivation for using hyperlink prediction is that adjacent nodes in a hyperedge are more likely to be mutually related. We phrase this property as mutual connectivity. 

Mutual connectivity highlights the importance of predicting adjacent nodes. We propose the hyperlink prediction task, where given a hyperedge $\hyperedge_h$, an anchor node $\node_i$ which is incident with $\hyperedge_h$, the positive node $\node_j$ which is also incident with $\hyperedge_h$ is discriminated from a negative node $\node_k$ which is not incident with $\hyperedge_h$. Furthermore, we corrupt the representation of the positive sample $\node_j$ to increase the difficulty of discrimination and the robustness of pre-training tasks.

Existing graph pre-training methods corrupt node representations by cutting down some edges and creating a subgraph, and then using a graph encoder on the subgraph~\cite{lu2021learning,liu2023graphprompt} to derive a representation of the target node. This strategy is infeasible in hypergraphs because hyperedges are indescomposable~\cite{tu2018structural}. If we cut down the adjacency between the positive node and the anchor node, the hyperedge is broken, the meaning of the hyperedge is incomplete, and high-order/multi-faceted information is discarded. 

To avoid disrupting the high-order/multi-faceted information, we propose to corrupt the positive sample by combining direct node perturbation (i.e., \nodepert) and incidence matrix perturbation (\incimatrixpert).  

\textbf{Node perturbation} randomly draws a node embedding $\embnode^g_j$ from: 
\begin{equation}\label{equ:np}
    \embnode^g_j = \mathcal{N}(\embnode_j,\lambda\mathbf{I}),
\end{equation}
where $\mathcal{N}$ is the Gaussian distribution with the original node representation as its mean, $\mathbf{I}$ is the identity matrix, and $\lambda$ is the covariance scalar. 

\textbf{Incidence matrix perturbation}. To avoid information leakage, i.e., the node embedding may contain information about the condition hyperedge $\hyperedge_h$, we remove the connection and rewrite the incidence matrix by letting $h(\node_j,\hyperedge_h)=0$. The corrupted incidence matrix is denoted as $\hat{\hgmatrix}$. We then propagate the filtered hyperedges back to nodes by:
\begin{equation}\label{equ:imp}
\embnode^a_j = (\hat{\mathbf{D}}^{v^{-1}}\hat{\hgmatrix}\mathbf{W}\hyperedgematrix)_j,
\end{equation}
where $(\cdot)_j$ represents the j-th row elementsm,$\hat{\mathbf{D}}^{v^{-1}}$ is the the degree matrix of nodes after the perturbation.


Then we merge $\embnode^g_i$ and $\embnode^a_i$ and obtain the corrupted node representation $\widetilde{\embnode}_j$, which is shown below:
\begin{equation}
\widetilde{\embnode}_j = \embnode^g_j+\embnode^a_j.
\end{equation}

Given the hyperedge $\hyperedge_h$, the anchor node $\node_i$, the positive node $\node_j$ and the negative node $\node_k$. We use the ranking loss $\ploss$ as the training objective of the hyperlink prediction task (\modelacro-P), which is shown below:
\begin{equation}
\label{equa:lploss}
    \ploss = -\sum_{i,j,k}\sigma(\embnode_i\cdot\widetilde{\embnode}_j-\embnode_i\cdot\embnode_k),
\end{equation}
where $\sigma(\cdot)$ is the sigmoid function.

\subsection{Hypergraph contrastive learning}\label{sec:contrasttask}
Furthermore, as label information is unavailable during the pre-training stage, we introduce Hypergraph Self-Contrastive Learning (\hgselfcl) to better the global information of the hypergraph and the local variability among nodes within a hyperedge.

First, we propose self-contrastive inter-hyperedges. 
Formally, given a node set $\nodeset$ in a batch, we treat any node embedding $\embnode_i$ as the anchor, the positive sample is the anchor's augmented embedding ${\embnode}_i^a$ from the augmented hypergraph $\hat{\hg}$ by Equation~\ref{equ:imp}, and negative samples are other node embeddings $\embnode_j$ in $\nodeset$. 
We denote the inter-hyperedge contrastive loss as $\interloss$.
\begin{equation}
        \interloss = \frac{1}{\|\nodeset\|}\sum_{\node_i \in \nodeset}-\log\frac{\exp{(sim(\embnode_i,\embnode_i^a)/\tau)}}{\sum_{\node_j \in \nodeset}\exp{(sim(\embnode_i,\embnode_j})/\tau)},
\end{equation}
where $\tau$ is a temperature parameter.

Second, we propose intra-hyperedge self-contrastive learning. Formally, given a hyperedge set $\hyperedgeset$ in a batch, we sample a node set $\nodeset_h$ from each hyperedge $\hyperedge_h\in\hyperedgeset$ and $\|\nodeset_h\|=K$. Given an anchor node $\node_i \in \nodeset_h$, its positive sample $\embnode_i^a$ is obtained by Equation~\ref{equ:imp}, and its negative samples are other nodes embedding $\node_j\in \hyperedge_h$. We denote the intra-
hyperedge contrastive loss as $\intraloss$.
\begin{small}
    \begin{equation}
    \intraloss = \frac{1}{\|\hyperedgeset\|\|\nodeset_h\|}\sum_{\hedges_h \in \hyperedgeset}\sum_{\nodes_i \in \nodeset_h}-\log\frac{\exp{(sim(\embnode_i,\embnode_i^a)/\tau))}}{\sum_{\nodes_j \in \nodeset_h}\exp{(sim(\embnode_i,\embnode_j)/\tau))}},
\end{equation}
\end{small}
where $\tau$ is a temperature parameter.

The overall loss in the pre-training stage includes the objective loss in hyperlink prediction $\ploss$ and the hypergraph self-contrastive from global and local levels, as shown below
\begin{equation}
\preloss = \ploss + \beta_1\intraloss+\beta_2\interloss.
\end{equation}
where $\beta_1$ and $\beta_2$ are hyper-parameters.

\section{Fine-tuning on User Feedback}
In recommendation systems, TOP-N recommendation and sequence recommendation are two of the most common recommendation tasks. $\modelacro$ designs a simple yet effective downstream fine-tuning approach for both tasks.

\subsection{Top-N recommendation}
In the pre-training phase, users and items are isolated, and there is no direct connection between them. Therefore, as shown in Figure~\ref{fig:framework}(b), in the fine-tuning stage, we first establish connections between users and items. 

\textbf{Hypergraph updating}. We update the incidence matrix according to the interaction between users and items. 
Formally, if user $\usernode_i$ is connected to hyperedge $\hyperedge_p$ and item $\itemnode_j$ is connected to hyperedge $\hyperedge_q$, and $\usernode_i$ interacts with $\itemnode_j$ (e.g., buy or click), we set $h(\usernode_i, \hyperedge_q) = 1$ and $h(\itemnode_j, \hyperedge_p) = 1$. 

The node embedding $\check{\nodematrix}$ is updated by:
\begin{equation}
\label{equa:hgupda}
\begin{aligned}
\check{\nodematrix} &= \nodematrix+\tilde{\mathbf{D}}^{v^{-1}}\tilde{\mathbf{H}}\mathbf{W}\hyperedgematrix.
\end{aligned}
\end{equation}
where $\hyperedgematrix$ is obtained in the pre-training stage and $\mathbf{W}$ is the parameters obtained in the pre-training stage, $\tilde{\mathbf{H}}$ is the updated hypergraph, $\tilde{\mathbf{D}}^{v^{-1}}$ is the degree matrix of nodes after the hypergraph updated.

\textbf{LightGCN fine-tuning}.
To comprehensively understand the relationship between users and items, we build a user-item bipartite graph $\uigraph$ and employ the classical LightGCN to finetune the representations
\begin{equation}
\ftembuser_i,\ftembitem_j = LightGCN(\check{\embuser}_i,\check{\embitem}_j,\uigraph).
\end{equation}
where $\check{\embuser}_i$, and $\check{\embitem}_j$ are the user embedding and the item embedding in $\check{\nodematrix}$, respectively. 

Then, we introduce $\overalltopnloss$ as the training objective in the fine-tuning stage
\begin{equation}
\overalltopnloss = -\sum_{i,j,k}\sigma(\ftembuser_i\ftembitem_j-\ftembuser_i\ftembitem_k).
\end{equation}
where the user $i$ interacts with item $j$ and not interacts with item $k$, $\sigma(\cdot)$ is the sigmoid function.


\subsection{Sequential recommendation}
Unlike the TOP-N recommendation, sequential recommendation focuses on not only long-term user intent but also short-term user intent to make item predictions for the next moment. We believe that LightGCN only captures the long-term user intent, therefore we extra employ GRU\cite{chung2014empirical} to capture short-term user intent.

Formally, given a user $i$ 's interaction sequence on most recent $s$ items,$\mathcal{V}_i = \{\itemnode_{t-s+1}, \cdots, \itemnode_{t}\}$, our model $\modelacro$ learns the user's short-term intents at moment $t$ by GRU\cite{chung2014empirical}
\begin{equation}
\begin{aligned}
    \mathbf{h}_i^t &= GRU(\mathcal{V}_i),\\
    \dot{\embuser}_i^t &= \overline{\embuser}_i+\mathbf{h}_i^t.\\
\end{aligned}
\end{equation}
where $\mathbf{h}_i^t$ is the hidden state of user $i$ at time $t$. The user representation$\dot{\embuser}_i^t$ at time $t$ is an ensemble of short-term user intent and long-term user preference.

Then, We introduce the sequential recommendation loss $\overallseqloss$.
\begin{equation}
    \overallseqloss = -\sum_{i,j,k}\sigma(\dot{\embuser}_i^t\ftembitem_j-\dot{\embuser}_i^t\ftembitem_k).
\end{equation}
where the user $i$ interacts with item $j$ and not interacts with item $k$ at time $t+1$, $\sigma(\cdot)$ is the sigmoid function.

\begin{table}[!t]
\caption{Statistics of the datasets}
\label{tab:dataset}
\centering
\resizebox{0.98\linewidth}{!}{
\centering
\begin{tabular}{l|r|r|r|r|r|r}
\hline
Dataset     & \#User & \#Item & \#Interactions & \#Relationship & \#Side Information & \#Density \\ \hline
Gowalla     & 10,671          & 48,587          & 1,195,532                         & 93,260             &Friendship,POI location            & 0.23\%             \\
Foursquare  & 22,065          & 9,891           & 182,044                           & 79,704            &Friendship,POI location             & 0.08\%             \\
Books & 28,898          & 85,188          & 3,233,028                         & 160,448,633              &Product review, category                & 0.13\%             \\ \hline
\end{tabular}}
\end{table}

\section{EXPERIMENT}
\label{sec:exp}
In this section, we conduct experiments to study the following questions:
\begin{enumerate}
\item[\textbf{RQ1:}] Can $\modelacro$ generalize well to different side information? 
\item[\textbf{RQ2:}] How does each component in $\modelacro$ contribute to the overall performance? 
\item[\textbf{RQ3:}] Is $\modelacro$ suitable for pre-training hypergraphs?
\item[\textbf{RQ4:}] Does $\modelacro$ alleviate the users and items cold start problems?
\end{enumerate}
\subsection{Experimental Setup}
\textbf{Datasets:}
Our experiments are conducted on three real-world datasets: Gowalla~\cite{yin2015modeling}, Foursquare~\cite{feng2018deepmove}, and Books\cite{mcauley2015image}. We evaluate \modelacro on these datasets because they are evaluated frequently and cover a wide range of common types of side information in recommendation systems, including item brand, item category, item review, user social networks, and POI geo-location. These datasets include timestamped feedback. Thus, they can be used for Top-N recommendations and sequential recommendations.  Statistics of the datasets are shown in Table~\ref{tab:dataset}.

\textbf{Evaluation Metrics}
We evaluated the TOP-N recommendation performance and Sequential recommendation performance of all models on three datasets using two widely-used metrics, N@K(NDCG@K) and R@K(Recall@K), where K=[10,20]. 
We adopt the widely used leave-one-out evaluation method, similar to SPEX~\cite{li2021spex}. 

\textbf{Competitors}. We compare \modelacro to an extensive list of competitors. We classify based on whether competitors use side information and pre-training.
For the TOP-N task, the competitors are (1).without side information and pre-training:  NGCF~\cite{wang2019neural}, LightGCN~\cite{he2020lightgcn},SGL~\cite{wu2021self}, HCCF~\cite{xia2022hypergraph},
(2). with pre-training but without side information: SimRec~\cite{xia2023graph},
(3). with side information but without pre-training: SPEX~\cite{li2021spex}, HIRE~\cite{liu2019recommender}, 
FREEDOM~\cite{zhou2022tale}, Flashback\cite{yang2020location},  HGNN$^+$~\cite{gao2022hgnn},
(4). with side information and pre-training: Graph-Flashback~\cite{rao2022graph}. 
There are three categories in (3): Feature~\cite{liu2019recommender}, Signal~\cite{li2021spex} and Model~\cite{zhou2022tale,yang2020location,/FengYZJG19HGNN,gao2022hgnn}.

For the Sequential task, the competitors are 
(1). without side information and pre-training:  SASRec\cite{kang2018self}, ContrastVAE\cite{wang2022contrastvae}, BRET4Rec\cite{sun2019bert4rec}, CBiT\cite{du2022contrastive},
(2). with side information but without pre-training:   Flashback\cite{yang2020location},HGNN$^+$~\cite{gao2022hgnn},
(3). with side information and pre-training: S$^3$-Rec\cite{zhou2020s3},Graph-Flashback\cite{rao2022graph}.
More details about competitors can be found in the related works~\ref{sec:related} section.

\textbf{Implementations}. The proposed $\modelacro$ framework is implemented in Pytorch library and Deep HyperGraph\cite{gao2022hgnn,feng2019hypergraph}. We adopt the Adam\cite{kinga2015method} optimizer in both the pre-training and fine-tuning stages. In the pre-training stage, the embedding sizes of the nodes and hyperedges are both 64, the batch size is 4096, the learning rate is 0.0005, and the epochs are 500. The noise intensity $\lambda$ is 0.1. The hypergraph contrastive learning hyperparameter $\beta_1$ and $\beta_2$ are 0.005 and 0.01, respectively. For the updated hypergraph, we only use the training dataset. In the fine-tuning stage, The epochs are 10, and the learning rate is 0.0005. During the first three epochs, we amplify the learning rate by ten times. The LightGCN layer number $K$ is 2. Models are in the same setting for fairness. The purpose is to expedite the transfer of the pre-trained universal representations to downstream tasks. To potentially address oversmoothing, we also explore the possibility of incorporating contrastive learning and denoising techniques\cite{zhou2022tale,yu2022graph}. 

\begin{table*}[!t]
\caption{Comparative performance for TOP-N recommendation, bold fonts for best results, underlined scores for the second best results. }
\label{tab:topnresult}
\resizebox{1\linewidth}{!}{
\centering
\renewcommand\arraystretch{1.5}
\begin{tabular}{l|rrrr|rrrr|rrrr}
\hline
\multicolumn{1}{c|}{\multirow{2}{*}{}} & \multicolumn{4}{c|}{Gowalla}                                                                                               & \multicolumn{4}{c|}{Foursquare}                                                                                            & \multicolumn{4}{c}{Book}                                                                                                  \\ \cline{2-13}
\multicolumn{1}{c|}{}                  & \multicolumn{1}{r}{N@10} & \multicolumn{1}{r}{N@20} & \multicolumn{1}{r}{R@10} & \multicolumn{1}{r|}{R@20} & \multicolumn{1}{r}{N@10} & \multicolumn{1}{r}{N@20} & \multicolumn{1}{r}{R@10} & \multicolumn{1}{r|}{R@20} & \multicolumn{1}{r}{N@10} & \multicolumn{1}{r}{N@20} & \multicolumn{1}{r}{R@10} & \multicolumn{1}{r}{R@20} \\ \hline
NGCF                                  & 0.5091                      & 0.5393                      & 0.7653                        & 0.8839                        & 0.5513                      & 0.5735                      & 0.6894                        & 0.7767                        & 0.6593                      & 0.6748                      & {\ul 0.8894}                  & {\ul 0.9501}                  \\
LightGCN                              & 0.2782                      & 0.3095                      & 0.4257                        & 0.5502                        & 0.6091                      & 0.6276                      & 0.7709                        & 0.8439                        & 0.5792                      & 0.5994                      & 0.8303                        & 0.9097                        \\

SGL                                   & 0.5607                      & 0.5875                      & 0.7501                        & 0.8566                        & 0.6192                      & 0.6336                      & 0.7597                        & 0.8164                        & {\ul 0.6689}                & {\ul 0.6845}                & 0.8433                        & 0.9047                        \\
HCCF                                  & 0.6080                      & 0.6281                      & 0.8624                        & 0.9406                        & 0.5678                      & 0.5882                      & 0.7204                        & 0.8011                        & 0.6005                      & 0.6200                      & 0.8550                        & 0.9318                        \\
SimRec                                & {\ul 0.6132}                & {\ul 0.6336}                & {\ul 0.8682}                  & {\ul 0.9478}                  & {\ul 0.6274}                & {\ul 0.6474}                & {\ul 0.7984}                  & {\ul 0.8772}                  & 0.6026                      & 0.6231                      & 0.8552                        & 0.9354                        \\ \hline
\makecell[l]{NGCF\\\&SPEX}                            & 0.3542                      & 0.3919                      & 0.5571                        & 0.7063                        & 0.5357                      & 0.5665                      & 0.6670                        & 0.7882                        & 0.5665                      & 0.5881                      & 0.8215                        & 0.9065                        \\
\makecell[l]{LightGCN\\\&SPEX}                        & 0.4469                      & 0.4810                      & 0.6802                        & 0.8147                        & 0.5929                      & 0.6119                      & 0.7350                        & 0.8278                        & 0.6573                      & 0.6723                      & 0.8801                        & 0.9403                        \\   \hline
HIRE                                  & 0.2335                      & 0.2662                      & 0.3825                        & 0.5120                        & 0.5679                      & 0.5889                      & 0.7249                        & 0.8081                        & 0.2887                      & 0.3365                      & 0.5056                        & 0.6952                        \\  \hline
FREEDOM                               & 0.5916                      & 0.6160                      & 0.8536                        & 0.9496                        & 0.6151                      & 0.6404                      & 0.7740                        & 0.8741                        & N/A                         & N/A                         & N/A                           & N/A                           \\
Flashback                             & 0.1354                      & 0.1663                      & 0.2427                        & 0.3655                        & 0.5278                      & 0.5490                      & 0.6819                        & 0.7663                        & N/A                         & N/A                         & N/A                           & N/A                           \\
HGNN$^+$                                 & 0.6256                      & 0.6630                      & 0.7244                        & 0.8711                        & 0.3859                      & 0.4289                      & 0.5013                        & 0.6699                        & 0.1914                      & 0.2241                      & 0.2904                        & 0.4207                        \\  \hline
\makecell[l]{Graph-\\Flashback}                       & 0.2052                      & 0.2376                      & 0.3384                        & 0.4672                        & 0.5398                      & 0.5621                      & 0.6999                        & 0.7879                        & N/A                         & N/A                         & N/A                           & N/A                           \\
GENET                                 & \textbf{0.7193}             & \textbf{0.7289}             & \textbf{0.9412}               & \textbf{0.9785}               & \textbf{0.7081}             & \textbf{0.7241}             & \textbf{0.8835}               & \textbf{0.9463}               & \textbf{0.7009}             & \textbf{0.7129}             & \textbf{0.9159}               & \textbf{0.9626}       \\  \hline     
\end{tabular}
}
\scriptsize{N/A indicates the model cannot be fitted into a Nvidia GeForce GTX 3090 GPU card with 24 GB memory. }
\end{table*}
\begin{table*}[!t]
\caption{Comparative performance on Sequential recommendation, bold fonts for best results, underlined scores for the second best results.}
\label{tab:seqresult}
\resizebox{1\linewidth}{!}{
\centering
\renewcommand\arraystretch{1.5} 
\begin{tabular}{l|rrrr|rrrr|rrrr}
\hline
         & \multicolumn{4}{c|}{Gowalla}                                                                                     & \multicolumn{4}{c|}{Foursquare}                                                                                            & \multicolumn{4}{c}{Book}                                                                                           \\ \cline{2-13}
          & \multicolumn{1}{r}{N@10} & \multicolumn{1}{r}{N@20} & \multicolumn{1}{r}{R@10} & \multicolumn{1}{r|}{R@20} & \multicolumn{1}{r}{N@10} & \multicolumn{1}{r}{N@20} & \multicolumn{1}{r}{R@10} & \multicolumn{1}{r|}{R@20} & \multicolumn{1}{r}{N@10} & \multicolumn{1}{r}{N@20} & \multicolumn{1}{r}{R@10} & \multicolumn{1}{r}{R@20} \\  \hline
SASRec          & 0.2516                      & 0.2810                      & 0.3784                   & 0.4954                   & 0.6425                      & 0.6605                      & 0.7906                        & 0.8620                        & 0.2474                      & 0.2904                      & 0.4489                        & 0.6192                        \\
BRET4Rec        & 0.2693                      & 0.3031                      & 0.4188                   & 0.5531                   & 0.6988                      & 0.7140                      & 0.8424                        & 0.9024                        & 0.4329                      & 0.4624                      & 0.6288                        & 0.7455                        \\
ContrastVAE     & 0.5403                      & 0.5680                      & 0.8129                   & 0.9211                   & 0.6926                      & 0.7095                      & 0.8472                        & 0.9138                        & 0.3868                      & 0.4234                      & 0.6295                        & 0.7734                        \\
CBiT            & 0.6591                &  0.6766                & 0.8592             & 0.9273             & 0.6460                & 0.6639                &0.8485                  &  0.9083                  & {\ul 0.4356}                & {\ul 0.4663}                & 0.6440                        & 0.7649                        \\ \hline
HGNN$^+$           & 0.6641                      & 0.6955                      & 0.7618                   & 0.8846                   & 0.2550                      & 0.3019                      & 0.3841                        & 0.5701                        & 0.2380                      & 0.2531                      & 0.3301                        & 0.3892                        \\
Flashback       & 0.7087                      & 0.7223                      & 0.8873                   & 0.9406                   & 0.7270                      & 0.7392                      & 0.8427                        & 0.8909                        & \multicolumn{1}{r}{N/A}     & \multicolumn{1}{r}{N/A}     & \multicolumn{1}{r}{N/A}       & \multicolumn{1}{r}{N/A}       \\ \hline
S$^3$-Rec              & 0.5819                      & 0.6021                      & 0.8308                   & 0.9097                   & 0.6634                      & 0.6801                      & 0.8241                        & 0.8897                        & 0.4239                      & 0.4573                      & {\ul 0.6738}                  & {\ul 0.8051}                  \\
\makecell[l]{Graph-\\Flashback} & {\ul 0.7282}                & {\ul 0.7408}                & {\ul 0.8939}             & {\ul 0.9432}             & {\ul 0.7368}                & {\ul 0.7479}                & {\ul 0.8508}                  & {\ul 0.8987}                  & \multicolumn{1}{r}{N/A}     & \multicolumn{1}{r}{N/A}     & \multicolumn{1}{r}{N/A}       & \multicolumn{1}{r}{N/A}       \\
GENET           & \textbf{0.7505}             & \textbf{0.7578}             & \textbf{0.9468}          & \textbf{0.9752}          & \textbf{0.7561}             & \textbf{0.7690}             & \textbf{0.9040}               & \textbf{0.9547}               & \textbf{0.6006}             & \textbf{0.6211}             & \textbf{0.8057}               & \textbf{0.8866}        \\ \hline      
\end{tabular}}
\scriptsize{N/A indicates the model cannot be fitted into a Nvidia GeForce GTX 3090 GPU card with 24 GB memory. }
\end{table*}
\subsection{Generalization of \modelacro}
To answer \textbf{RQ1}, we evaluate $\modelacro$ on three datasets (Gowalla, Foursquare, Books) for two recommendation tasks (Top-N and sequential recommendation), showing superior performance across all metrics and tasks. For instance, in Top-N recommendation, it increased NDCG@10 by up to $14.98\%$ and Recall@10 by up to $10.66\%$, shown in Table~\ref{tab:topnresult}. In sequential recommendation, the increase was up to $37.88\%$ for NDCG@10 and $19.58\%$ for Recall@10, shown in Table~\ref{tab:seqresult}. Existing methods demonstrated weaker generalization, especially across different datasets. $\modelacro$ consistently achieved a recommendation performance of over $0.6000$ for NDCG@10 across all datasets and tasks, indicating strong generalization capability. Moreover, pre-training positively impacted recommendation performance. Models like SimRec and Graph-Flashback significantly improved after pre-training. S$^3$-rec, by incorporating pre-training into SASRec, achieved improvements of 119.56\% and 131.28\% in Recall@10 and NDCG@10, respectively, on the Gowalla dataset. Overall, $\modelacro$ demonstrated robust generalization across three datasets and two tasks, outperforming 17 competitors in various side information scenarios.

\subsection{Ablation Study}
To answer \textbf{RQ2}, we assessed the effectiveness of various variants of \modelacro by removing components such as the node perturbation method $\nodepert$, incidence matrix perturbation $\incimatrixpert$, and hypergraph self-contrastive learning $\hgselfcl$ in pre-training. As is shown in Figure ~\ref{fig:otherexp}(a), the results indicated that each component is crucial for performance. On the Gowalla dataset, removing $\nodepert$(i.e.,"w/o" $\nodepert$), $\incimatrixpert$(i.e.,"w/o" $\incimatrixpert$), and $\hgselfcl$(i.e.,"w/o" $\hgselfcl$) led to declines in Recall@10 and NDCG@10 by $0.0124$ and $0.0185$, $0.0181$ and $0.0332$, $0.0103$ and $0.0196$, respectively. Particularly, removing $\incimatrixpert$ had the most significant impact as it is a core component of the pre-training stage. Its removal left only $\nodepert$ for node perturbation, oversimplifying the link prediction task and affecting the model's ability to learn intrinsic relationships and higher-order dependencies within the data.

\subsection{Pre-training on Hypergraphs}
To study whether the proposed pre-training tasks are suitable for pre-training on hypergraphs (\textbf{RQ3}), we studied the suitability of pre-training tasks for hypergraphs, comparing the pre-training phase (\modelacro-P) of \modelacro with two common graph-based pretext tasks: Link Prediction (LP) and Node Feature Reconstruction (NR). We utilized two popular graph pre-training strategies: Contrastive and Generative. Consequently, we obtained four pre-training tasks: LP-C, LP-G, NR-C, and NR-G. As a baseline, we also implemented random initialization of node embeddings. These pre-training methods were evaluated by directly computing user-item similarity for recommendation (No-tuning) and performance after fine-tuning with downstream models, including Recall@10 and NDCG@10, which are shown in ~\ref{fig:otherexp}(b)(c).

Based on the results, we draw the following conclusions: (1) \modelacro-P outperforms all other pre-training tasks on hypergraphs. For instance, on the FourSquare dataset, it increases Recall@10 by 28.24\% and NDCG@1 by 30.86\% compared to the best competitor (i.e., LP-C). (2) \modelacro-P is already very powerful without fine-tuning on feedback data, especially on the BOOK dataset, where all competitors' Recall@10 is under 0.5379 (0.2792 in NDCG@10), and \modelacro-P's performance achieves 0.8946(0.6700 in NDCG@10). (3) Compared to not pre-training and directly training the downstream model (Random), all pre-training strategies lead to significant performance improvement. (4) The link prediction (LP) pre-training strategy performs better than the node feature reconstruction (NR) strategy. (5) The contrastive strategy is superior to the generative strategy in pre-training generative strategy, resulting in lower uncertainty.

\subsection{Cold-start}
To answer \textbf{RQ4}, we evaluate the performance of $\modelacro$ in scenarios where it encounters user cold-start and item cold-start situations. 

To create a user cold-start scenario, we select the last 1\% of users based on their interaction counts and designate them as cold-start users. We remove these cold-start users from the training dataset and retain only cold-start users in the test dataset. We create an item cold-start scenario in the same manner.

\begin{figure*}[!t]
	\centering
\includegraphics[width=1\linewidth]{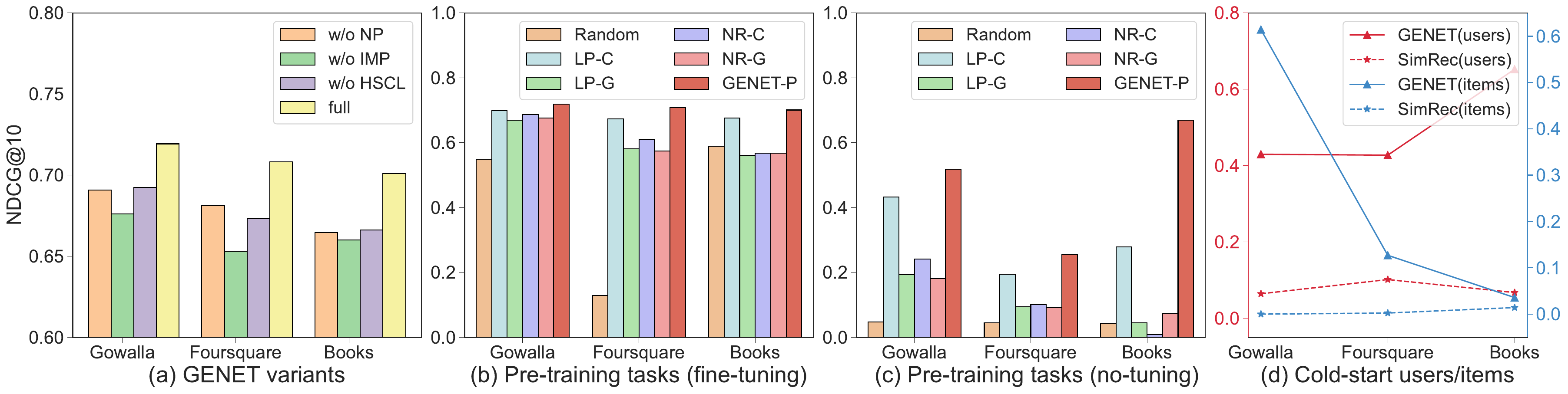}
	\caption{Performances of GENET variants, different pre-training tasks and cold start.}
	\label{fig:otherexp}
\end{figure*}

As shown in Figure~\ref{fig:otherexp}(d), $\modelacro$ demonstrates a remarkable capability to deal with cold-start users and items. It has a 321.50\% to 869.53\% improvement in NDCG@10 and a 258.98\% to 538.89\% improvement in Recall@10 compared with SimRec in the users cold-start scenario, and it has a 33.33\% to 4069.35\% improvement in Recall@10 and 152.11\% to 5191.67\% improvement in NDCG@10 in the items cold-start scenario. This highlights the effectiveness of the hypergraph constructed from social networks and user reviews in accurately modeling users' preferences and the effectiveness of the hypergraph constructed from POI location and item features in accurately modeling items' properties.

\section{CONCLUSION}
In this work, we proposed a novel framework called Generalized hypErgraph pretraiNing on sidE
informaTion(\modelacro). It aims to enhance recommendation performances by integrating side information. \modelacro is based on pretraining-finetuning, where heterogeneous side information is constructed as a unified hypergraph. We propose novel pre-training tasks tailored explicitly for hypergraphs, which can effectively capture high-order and multi-faceted relationships in hypergraphs. The finetuning is conducted with simple and straightforward approaches. Extensive experimental results show that \modelacro has excellent generalization. It outperforms SOTA competitors on two recommendation tasks on three public datasets with varying side information.

%
%
%
%

\bibliographystyle{splncs04}
\bibliography{ref}
\end{document}